\documentclass[conference]{IEEEtran}
\IEEEoverridecommandlockouts
\usepackage{cite}
\usepackage{amsmath,amssymb,amsfonts}
\usepackage{algorithmic}
\usepackage{graphicx}
\usepackage{textcomp}
\usepackage{xcolor}
\usepackage{amsmath,amssymb}
\usepackage{mathtools,leftindex}
\usepackage{makecell}

\usepackage{subcaption}
\usepackage{psfrag}
\usepackage{enumerate}
\usepackage{algorithmic}
\usepackage{algorithm}
\usepackage{acronym}
\usepackage{color, soul}
\usepackage{pifont}
\usepackage{bm}
\usepackage{afterpage}
\usepackage{multicol}
\usepackage{multirow}
\usepackage{array}
\usepackage[T1]{fontenc}
\usepackage[pscoord]{eso-pic}

\def\BibTeX{{\rm B\kern-.05em{\sc i\kern-.025em b}\kern-.08em
		T\kern-.1667em\lower.7ex\hbox{E}\kern-.125emX}}

\def\BibTeX{{\rm B\kern-.05em{\sc i\kern-.025em b}\kern-.08em
    T\kern-.1667em\lower.7ex\hbox{E}\kern-.125emX}}

\begin{document}
	
	\makeatletter
	\let\old@ps@headings\ps@headings
	\let\old@ps@IEEEtitlepagestyle\ps@IEEEtitlepagestyle
	\def\confheader#1{%
		\def\ps@IEEEtitlepagestyle{
			\old@ps@IEEEtitlepagestyle
			\def\@oddhead{\strut\hfill#1\hfill\strut}
			\def\@evenhead{\strut\hfill#1\hfill\strut}
		}
		\ps@headings
	}
	\makeatother
	\confheader{
		\small{Proceedings of the 13th RSI International Conference on Robotics and Mechatronics (ICRoM 2025), Dec. 16-18, 2025, Tehran, Iran} 
	}
	
	\newcommand{\placetextbox}[3]{
		\setbox0=\hbox{#3}
		\AddToShipoutPictureFG*{ \put(\LenToUnit{#1\paperwidth},\LenToUnit{#2\paperheight}){\vtop{{\null}\makebox[0pt][c]{#3}}}
		}
	}
	\placetextbox{.5}{0.055}{\textbf{\small{Proceedings of the 13th RSI International Conference on Robotics and Mechatronics (ICRoM 2025), Dec. 16-18, 2025, Tehran, Iran}}}
	
\acrodef{UAV}{Unmanned Aerial Vehicle}
\acrodef{ESC}{Electronic Speed Control}
\acrodef{BLDC}{Brushless Direct Current}
\acrodef{PCF}{Power Consumption Factor}
\acrodef{DoF}{Degrees of Freedom}
\acrodef{PD}{Proportional Derivative}
\acrodef{PID}{Proportional Integral Derivative}
\acrodef{VRS}{Vortex Ring State}
\acrodef{CG}{Center of Gravity}
\acrodef{SMC}{Sliding Mode Controller}
\acrodef{LSTM}{Long Short-Term Memory}
\acrodef{MAE}{Mean Absolute Error}
\acrodef{MLP}{Multi Layer Perceptron}
\acrodef{CNN}{Convolutional Neural Network}
\acrodef{PDP}{Plant Dynamics Predictor}

\title{Hybrid Neural Network and Conventional Controller Approach for Robust Control of Highly Unstable Systems: Application to Tilt-Rotor Control 
}

\author{\IEEEauthorblockN{ 1\textsuperscript{st} Ali Kafili Gavgani }
	\IEEEauthorblockA{
	 ali.kafili79@sharif.edu 
		\\
	 0009-0001-2763-4412 
		}
	\and
	\IEEEauthorblockN{ 2\textsuperscript{nd} Amin Talaeizadeh }
	\IEEEauthorblockA{
	amin.talaeizadeh@sharif.edu  \\
	 0000-0002-6997-7595}
	\and
	\IEEEauthorblockN{3\textsuperscript{th} Aria Alasty}
	\IEEEauthorblockA{aalasti@sharif.edu \\
0000-0002-6354-0034}
	\and
	\IEEEauthorblockN{4\textsuperscript{th} Hossein Nejat Pishkenari}
	\IEEEauthorblockA{
		nejat@sharif.edu  \\
		0000-0002-3487-3198
	}
	\and
	\;\;\;\;\;\;\;\;\;\;\;\;\;\;\;\;\;\;\; \textit{Advanced Research Lab for Control and Agricultural Robotics (Sharif AgRoLab)} \\
	\;\;\;\;\;\;\;\;\;\;\;\;\;\;\;\;\;\;\; \textit{Department of Mechanical Engineering, Sharif University of Technology, Tehran, Iran}
	
}

\maketitle

\begin{abstract}
Multirotors are widely used in applications ranging from surveillance to precision agriculture, yet conventional designs remain limited by their under-actuation. Tilt-rotor configurations overcome this limitation by enabling full actuation. This paper investigates neural-network-based control strategies for a fully actuated tilt-rotor system with four thrust-vectoring inputs.	 	
Our work is structured in two parts. First, we deliberately present a negative result by evaluating a direct input\textendash output control approach. In this method, multilayer perceptrons (MLPs), long short-term memory (LSTM) networks, and transformer models are trained to map system states and their desired values directly to control signals. We show that this strategy fails to stabilize the system, highlighting the inherent difficulty of applying direct input\textendash output learning to highly unstable plants.	 	
Second, as the main contribution, we propose a neural-network-enhanced sliding mode controller (SMC). The method decomposes the system dynamics into input-independent and input-dependent components, with the former learned from a small dataset using lightweight networks, thereby reducing real-time computational demands. Moreover, the proposed method can be trained using flight logs collected from low-performance controllers, and the resulting dynamic model learned from real-world data can be used in simulation.
We further compare MLP- and LSTM-based implementations under model uncertainties and external disturbances, demonstrating the robustness and effectiveness of the proposed approach; in particular, the controller with the LSTM plant dynamics predictor achieves superior performance to its MLP-based counterpart while also exhibiting lower runtime.
\end{abstract}


	\label{intro}

The rapid growth of multirotor applications \cite{surveillance, surveilanceRA, surveilanceRA2, package, PackDel, GavganiAgri, Agriculture2, shahrouz, search, nonedestructive, NDTRA1}, coupled with advances in deep learning, has encouraged researchers to integrate artificial neural networks into quadrotor control. Neural networks have been combined with diverse control strategies, such as PID \cite{AIPID,Arshia}, sliding mode control (SMC) \cite{AISMC}, model predictive control (MPC) \cite{AIMPC}, model reference adaptive control (MRAC) \cite{AIMRAC}, fuzzy-PID \cite{AIFUZZY}, and adaptive control schemes \cite{AIADAPT}.

At the same time, the under-actuation of conventional multirotors has motivated research on tilt-rotor designs, where rotors are tilted to achieve full actuation. Although these systems offer increased maneuverability, they also introduce additional complexity, and most studies have relied on conventional controllers to manage it \cite{ryll,zheng, tilthedralsegui,allenspach,KafiliOman,voliro}.

In this paper, we investigate a tilt-rotor system with four thrust-vectoring inputs, allowing independent control over all degrees of freedom (DoF). A conceptual design is shown in Fig. \ref{fig:conceptual}. Our work is structured into two parts:

\begin{figure}
	\centering
	\includegraphics[width=0.95\linewidth]{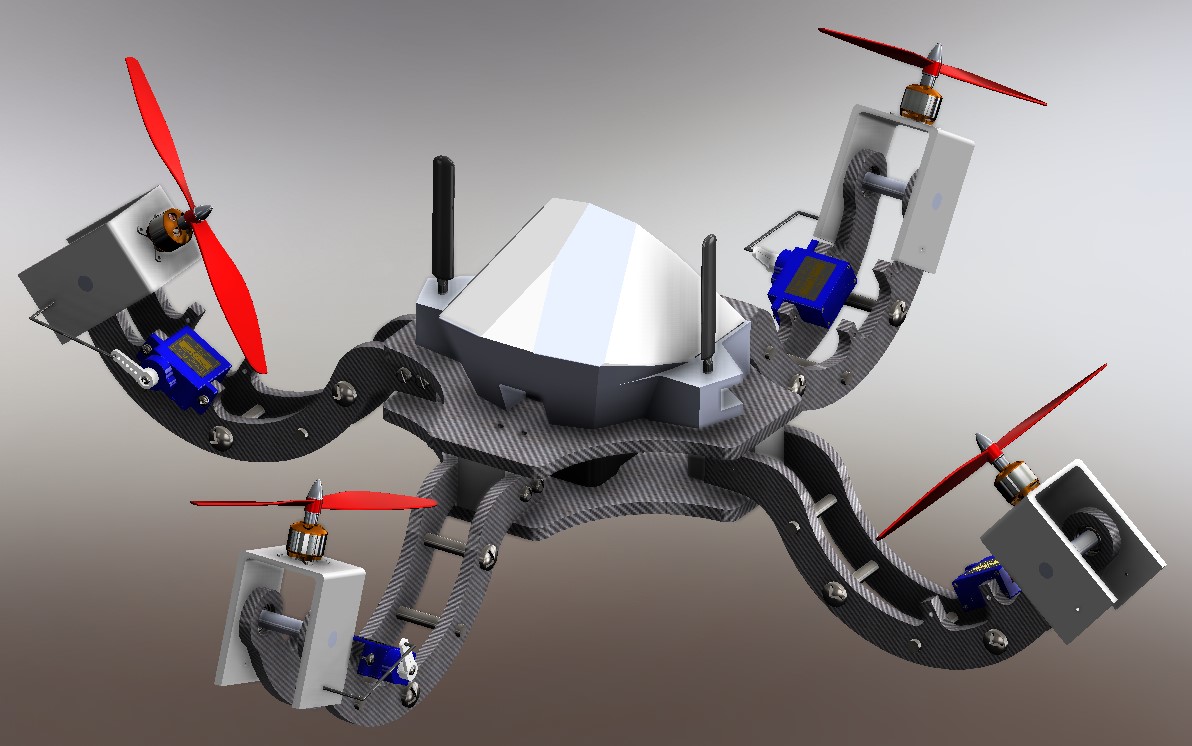}
	\caption{A conceptual design of the introduced tilt-rotor.} 
	\label{fig:conceptual}
\end{figure}

\textbf{First}, we present a deliberately negative result by attempting to control the system using a direct input\textendash output learning strategy. In this approach, multilayer perceptrons (MLPs), long short-term memory (LSTM) networks, and transformer models are trained to map the system states and their desired values to control inputs. We show that this method fails to stabilize the highly unstable tilt-rotor dynamics, illustrating the fundamental limitations of direct input\textendash output learning for such complex systems. While unsuccessful, these results serve an important role: they highlight the challenges involved and may still inspire further work on more stable platforms.

\textbf{Second}, as the main contribution of this paper, we introduce a neural-network-enhanced sliding mode control (SMC) method to robustly control the fully actuated tilt-rotor drone. In this approach, the system dynamics are decomposed into input-independent and input-dependent components, with the former learned efficiently from a small dataset using lightweight networks. This design lowers real-time computational requirements and enables deployment on low-cost microcontrollers. Embedding this learned model within the SMC framework ensures reliable and generalized control of the unstable system. Furthermore, we compare the performance of MLP- and LSTM-based controllers within this framework under uncertainties and external disturbances, demonstrating the robustness and effectiveness of the proposed method.
An additional advantage of this method is its ability to train on flight logs collected with low-performance, attitude-only controllers, as it focuses on learning plant parameters rather than control actions. Furthermore, this scheme can provide a dynamic model derived from real flights, which is useful for accurate simulations.

\section{Dynamic Modeling}
\label{sec:dyn}
\begin{figure}
	\centering
	\includegraphics[width=0.95\linewidth]{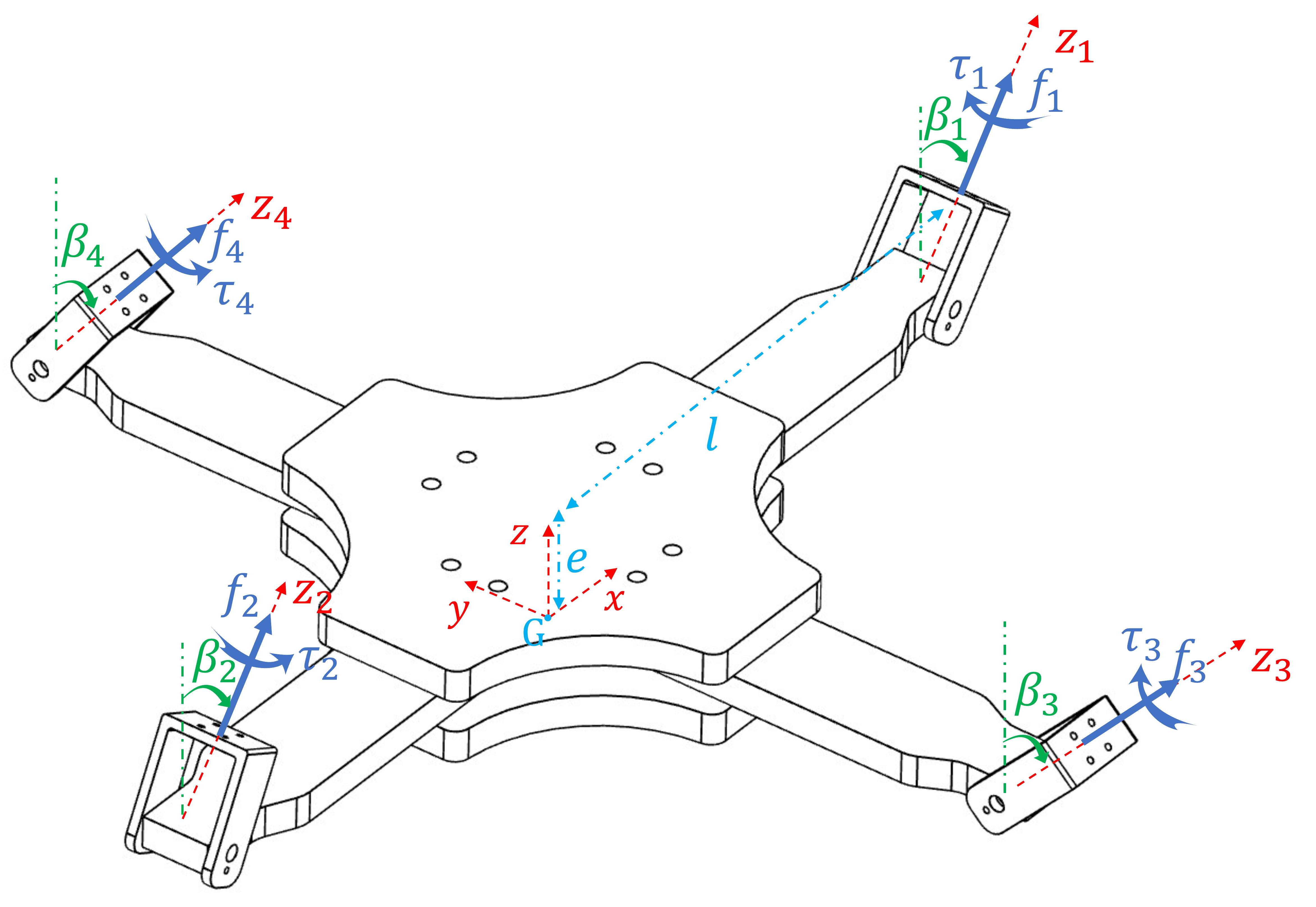}
	\caption{A simplified visualization of the system, emphasizing the free body diagram and the physical parameters.} 
	\label{fig:FBD}
\end{figure}

Here, we derive a mathematical model for system which is vital for our simulations. To have a manageable article length, we briefly discuss this section. A study by Gavgani et.al \cite{gavgani2025nextgenerationaerialrobots} can be referred for more details in dynamic modeling of the system. The coordinate systems and physical parameters are shown in \ref{fig:FBD}. The thrust forces $f_n=k_f \omega_n^2$ and the drag torques $\tau_n=k_m\omega_n^2$ , $n=1,2,3,4$ are executed by rotation of the propellers where $k_f$ and $k_m$ denote the aerodynamic characteristics of the propellers. $\omega_n$ is the angular velocity of the impeller $n$ and $\beta_n$ represents the hedral angle of the rotor axis $n$. $m$ indicates the mass of the body. The general form of Newton-Euler equation is as following:

\begin{equation}
	m \leftindex_I{\ddot{ \bm{p}}}= \leftindex_I{\bm{f}} + m \leftindex_I{\bm{g}} ,
\end{equation}
\begin{equation}
	\bm{J} \leftindex_B{\dot{ \bm{\omega}}}= -\leftindex_B{\bm{\Omega}} \bm{J} \leftindex_B{\bm{\omega}} +  \leftindex_B{\bm{\tau}} ,
\end{equation}
where \( I \) and \( B \) indices respectively denote the inertia and body frames. The rotation sequence from the inertia to the body frame involves the Euler angles \(\psi\), \(\theta\), and \(\phi\), corresponding to rotations about the \( Z \), \( y' \), and \( x'' \) axes in that order. \(\leftindex_I{\ddot{ \bm{p}}} = [\ddot{X} \quad \ddot{Y} \quad \ddot{Z}]^T\) is the acceleration vector in the inertia frame. $\bm{f}$ and $\bm{\tau}$ respectively represent the thrust and torque vectors produced by rotation of propellers. $\bm{g}$ is the gravity constant vector and $\bm{J}$ denotes the inertial moment matrix expressed in the body frame. $\bm{\omega}$ indicates the rotational velocity of the body and $\bm{\Omega}$ represents the skew-symmetric matrix of the $\bm{\omega}$.

\section{Sliding Mode Controller Design}
\label{sec:control}
Let
\begin{equation}
	\label{xbar}
	\bm{\bar{x}}= \begin{array}{cccccc}
		[X & Y & Z & \phi & \theta & \psi]^\intercal \end{array} ,
\end{equation}

Control input vector for Hedral quadrotor including the inputs for \ac{BLDC} and servo motors is expressed as
\begin{equation}
	\label{u}
	\begin{aligned}
		\bm{u}=[\begin{array}{l l l l l l l l}
			\omega_1 & \omega_2 & \omega_3 & \omega_4 & \beta_1 & \beta_2 	& \beta_3 & \beta_4 \\
		\end{array}]^\intercal .
	\end{aligned}
\end{equation}

The dynamic equations for Hedral quadrotor are not affine in input, causing the problem with the allocation of control signals to the actuators directly. To solve this problem, \cite{KafiliOman} offered the use of a virtual intermediary input set $\bm{v}$ to make the equations input-affine.

\begin{equation}
	\label{v}
	\bm{v}=\left[\begin{array}{cccc}
		v_1 & v_2 & ... & v_8 \end{array}\right]^\intercal,
\end{equation}

where
\begin{equation}
	\label{v2}
	v_i=\omega_i^2 \sin \beta_i ,\quad
	v_{i+4}= \omega_i^2 \cos \beta_i ,\quad i=1,2,3,4 .
\end{equation}

Thus the equations of motion for Hedral quadrotor can be described in the input-affine form
\begin{equation}
	\ddot{{\bar{\bm{x}}}} =\bm{a}+\bm{G}\bm{v} ,
\end{equation}

where
\begin{equation}
	\bm{G}= \frac{\partial \ddot{\bm{\bar{x}}}}{\partial \bm{v}} \quad , \quad \bm{a}=\ddot{\bm{\bar{x}}}-\frac{\partial \ddot{\bm{\bar{x}}} }{\partial \bm{v}} \bm{v} .
\end{equation}

Defining the sliding surfaces as
\begin{equation}
	\label{eq:sliding surfaces}
	\bm{s}=\bm{\dot{\bar{x}}}-\bm{\dot{\bar{x}}}_{d} + \bm{\lambda} (\bm{\bar{x}} - \bm{\bar{x}}_{d}) ,
\end{equation}

where $\bm{\bar{x}}_{d}$ represents the desired state of $\bm{\bar{x}}$ 
and $\bm{\lambda}$ is a $6\times6$ diagonal matrix with positive constant values on its diagonal \cite{slotine}.

The asymptotic stability condition of the system is
\begin{equation}
	\label{sliding cond}
	\frac{1}{2}\frac{d}{dt}s_i^2 \leq -\eta_i|s_i| \quad i=1,2,3,4 \quad ,
\end{equation}

where $s_i$ corresponds to the i-th element of $\bm{s}$ and $\eta_i$ denotes a strictly positive constant \cite{slotine}. The control law is formulated by

\begin{equation}
	\bm{G} \bm{v}=  \ddot{\bar{\bm{x}}}_{d}+\bm{\lambda} (\dot{\bar{\bm{x}}}_{d}- \dot{\bar{\bm{x}}} ) -\bm{a} -\bm{K} \tanh{(\sigma\bm{s})}=\bm{w} .
	\label{Gv}
\end{equation}

$\bm{G}$ is a none-square matrix, disabling directly calculation of $\bm{G}^{-1}$  to obtain $\bm{v}$. The equation (\ref{psudo}) is used to solve the system $\bm{G}_{6\times 8} \bm{v}_{8 \times 1}=\bm{w}_{6 \times 1}$. This approach extracts the minimum-normed solution from the infinite set of solutions, minimizing the power consumption considering (\ref{normv}) \cite{KafiliOman}.

\begin{equation}
	\label{psudo}
	\bm{v}=\bm{G}^\intercal(\bm{G} \bm{G}^\intercal )^{-1} \bm{w} .
\end{equation}

Norm of $\bm{v}$ is derived as
\begin{equation}
	\label{normv}
	\|\bm{v}\|=(v_1^2+v_2^2+...+v_8^2)^{1/2}=(\omega_1^4+\omega_2^4+\omega_3^4+\omega_4^4)^{1/2} .
\end{equation}


Finally, the primary control signals for actuators would be obtained as
\begin{equation}
	\label{ucomp}
	\begin{aligned}
		& \beta_i=\tan^{-1}(\frac{v_i}{v_{i+4}}), \\		
		& \omega_i=(v_i^2+v_{i+4}^2)^{1/4} \quad , i=1,2,3,4.			
	\end{aligned}
\end{equation}

\section{Direct Input-Output Training Method for Control}
\label{sec:failed}
In this section, we train several networks to replicate the behavior of the controller we designed. Inputs to the network are \(\bar{\bm{x}}\) as illustrated by (\ref{xbar}) and its desired vector \(\bar{\bm{x}}_d\) for each time step. The outputs of the network are the control signals for the next step, denoted by \(\bm{u}\) as represented in (\ref{u}).
Given the multi-level control allocation required for a conventional controller, the direct input-output control method may offer the advantage of reduced computational cost. 

We aim to investigate three key aspects:

\begin{itemize}
	\item Whether such a system can be properly trained to achieve acceptable test results, even without application to the dynamic system.
	\item Whether this methodology is applicable to such an unstable and intricate dynamic system.
	\item Whether this controller has a lower computational cost compared to a conventional one.
\end{itemize}

\subsection{Data Collection}
The dataset should cover a large combination of input values with as much as possible small number of samples. To achieve this, we set sinusoidal trajectories with different frequencies. This makes various combination of values for the elements of input vector. Also, five different maneuvers are taken into account, making the trajectories discontinuous to include some hard initial values in the dataset, along with refining the data with more various samples. The system with the controller designed in section  \ref{sec:control} is simulated for 200 seconds for each maneuver with a time step of 0.1 seconds, resulting in 10,000 samples. The maneuvers are represented by (\ref{sinus}) and detailed in Table \ref{table:maneuvers}.

\begin{equation}
	\begin{aligned}
		& trajectory_i=A_i \sin \nu_i t \\
		&	for \quad i=X,Y,Z, \phi,\theta, \psi
		\label{sinus}
	\end{aligned}
\end{equation}

\begin{table}[h!]
	\centering
	\caption{Amplitudes and frequencies of the sinusoidal maneuvers.}
	\begin{tabular}{|c|c|c|c|c|c|c|c|}
		\hline
		man. & \makecell{$A_X$ \\ $\nu_X$} & \makecell{$A_Y$ \\ $\nu_Y$} & \makecell{$A_Z$ \\ $\nu_Z$} & \makecell{$A_\phi$ \\ $\nu_\phi$} & \makecell{$A_\theta$ \\ $\nu_\theta$} & \makecell{$A_\psi$ \\ $\nu_\psi$} & \makecell{usage \\ samples} \\
		\hline
		1 & \makecell{3 \\ 0.3} & \makecell{3 \\ 0.4} & \makecell{3 \\ 0.1} & \makecell{$30^\circ$ \\ 0.4} & \makecell{$30^\circ$ \\ 0.3} & \makecell{$360^\circ$ \\ 0.2} & \makecell{$train$ \\ $2000$} \\
		\hline
		2 & \makecell{3 \\ 0.4} & \makecell{4 \\ 0.35} & \makecell{5 \\ 0.25} & \makecell{$25^\circ$ \\ 0.3} & \makecell{$30^\circ$ \\ 0.2} & \makecell{$360^\circ$ \\ 0.1} & \makecell{$train$ \\ $5000$} \\
		\hline
		3 & \makecell{5 \\ 0.3} & \makecell{5 \\ 0.2} & \makecell{5 \\ 0.1} & \makecell{$25^\circ$ \\ 0.15} & \makecell{$25^\circ$ \\ 0.35} & \makecell{$360^\circ$ \\ 0.15} & \makecell{$train$ \\ $1000$} \\
		\hline
		4 & \makecell{4 \\ 0.4} & \makecell{3.5 \\ 0.3} & \makecell{4 \\ 0.15} & \makecell{$30^\circ$ \\ 0.4} & \makecell{$30^\circ$ \\ 0.2} & \makecell{$360^\circ$ \\ 0.3} & \makecell{$validation$ \\ $1000$} \\
		\hline
		5 & \makecell{3 \\ 0.5} & \makecell{3 \\ 0.3} & \makecell{4 \\ 0.2} & \makecell{$30^\circ$ \\ 0.2} & \makecell{$30^\circ$ \\ 0.4} & \makecell{$360^\circ$ \\ 0.2}& \makecell{$test$ \\ $1000$} \\
		\hline
	\end{tabular}
	\label{table:maneuvers}
\end{table}

\subsection{MLP Network}
\begin{figure}[h!]
	\centering
	\includegraphics[width=0.8\linewidth]{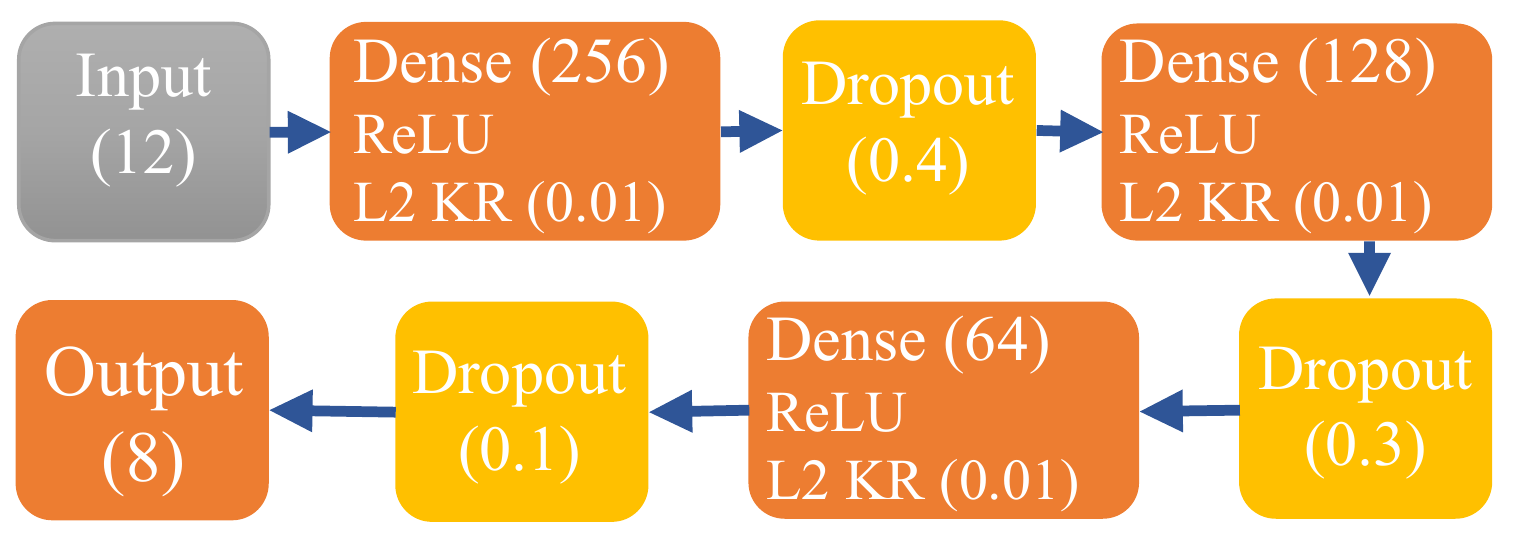}
	\caption{Structure of MLP network for direct input-output control method. (*KR: Kernel Regularizer)} 
	\label{fig: mlp_struct}
\end{figure}

A \ac{MLP} network with the architecture represented in Fig. \ref{fig: mlp_struct} is incorporated. ADAM optimizer is used in training process for 1400 epochs. This network exhibited 3.86 and 10.85 \ac{MAE} in train and test respectively.

\subsection{LSTM-Based Network}
\begin{figure}[h!]
	\centering
	\includegraphics[width=0.8\linewidth]{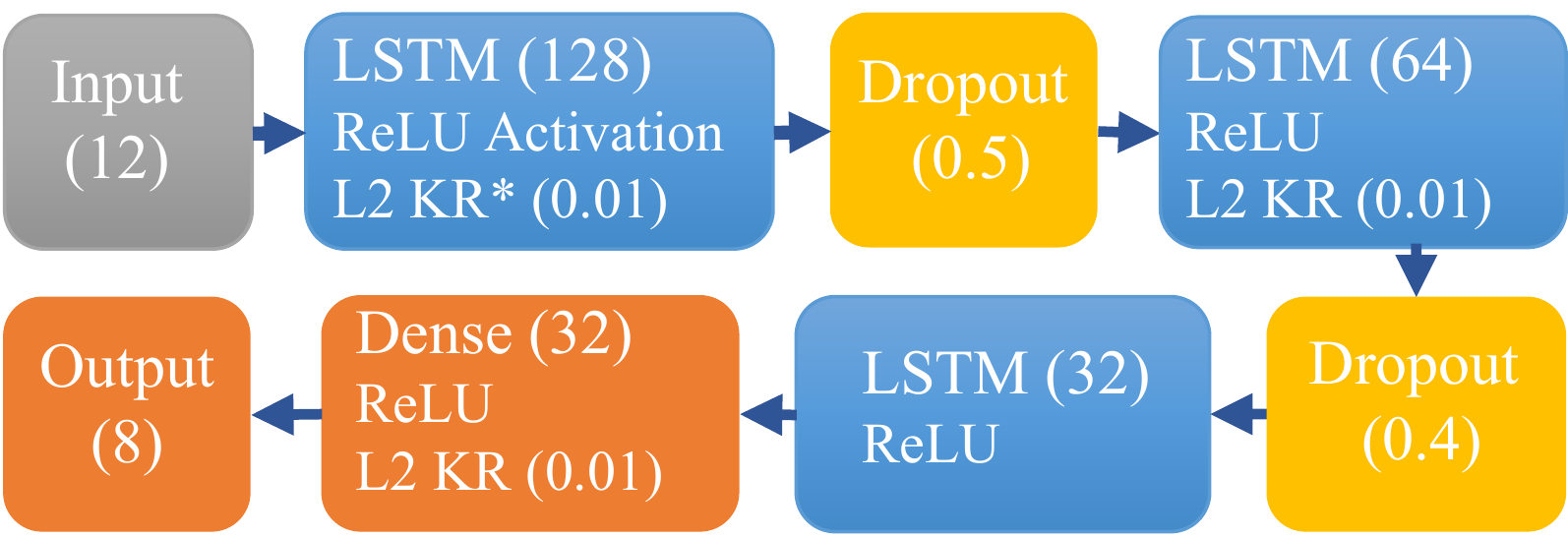}
	\caption{Structure of LSTM-based network for direct input-output control method. (*KR: Kernel Regularizer)} 
	\label{fig: lstm_struct}
\end{figure}

The architecture of the \ac{LSTM}-based network is visualized in Fig. \ref{fig: lstm_struct}. The network is trained using ADAM optimizer in 300 epochs. \ac{MAE} metric for train and test equals respectively to 2.62 and 10.61.

\subsection{Transformer-Based Network}
The transformer network is well-designed for capturing long dependencies, making it ideal for applications like text processing. Although dynamic systems typically exhibit shorter dependencies, we explored the applicability of the transformer network in training our system.

The transformer model is designed with the following layers:
\begin{itemize}
	\item Input Layer: Shape $(1, 12)$
	\item Transformer Encoder Layer:
	\begin{itemize}
		\item Normalization and Multi-Head Attention:
		\begin{itemize}
			\item Head Size: 64
			\item Number of Heads: 4
		\end{itemize}
		\item Feed Forward Dimension: 4
		\item Dropout: 0.3
		\item Feed Forward Network:
		\begin{itemize}
			\item Dense Layer with ReLU activation and 4 units
			\item Dropout: 0.3
			\item Dense Layer to match the input shape
		\end{itemize}
	\end{itemize}
	\item Flatten Layer
	\item Dense Layers:
	\begin{itemize}
		\item Dense Layer with 64 units and ReLU activation
		\item Dropout: 0.05
	\end{itemize}
	\item Output Layer: Dense Layer with 8 units
\end{itemize}

The network is trained using ADAM optimizer in 10000 epochs. \ac{MAE} metric for train and test equals respectively to 5.26 and 9.49.

\begin{figure}
	\centering
	\includegraphics[width=0.95\linewidth]{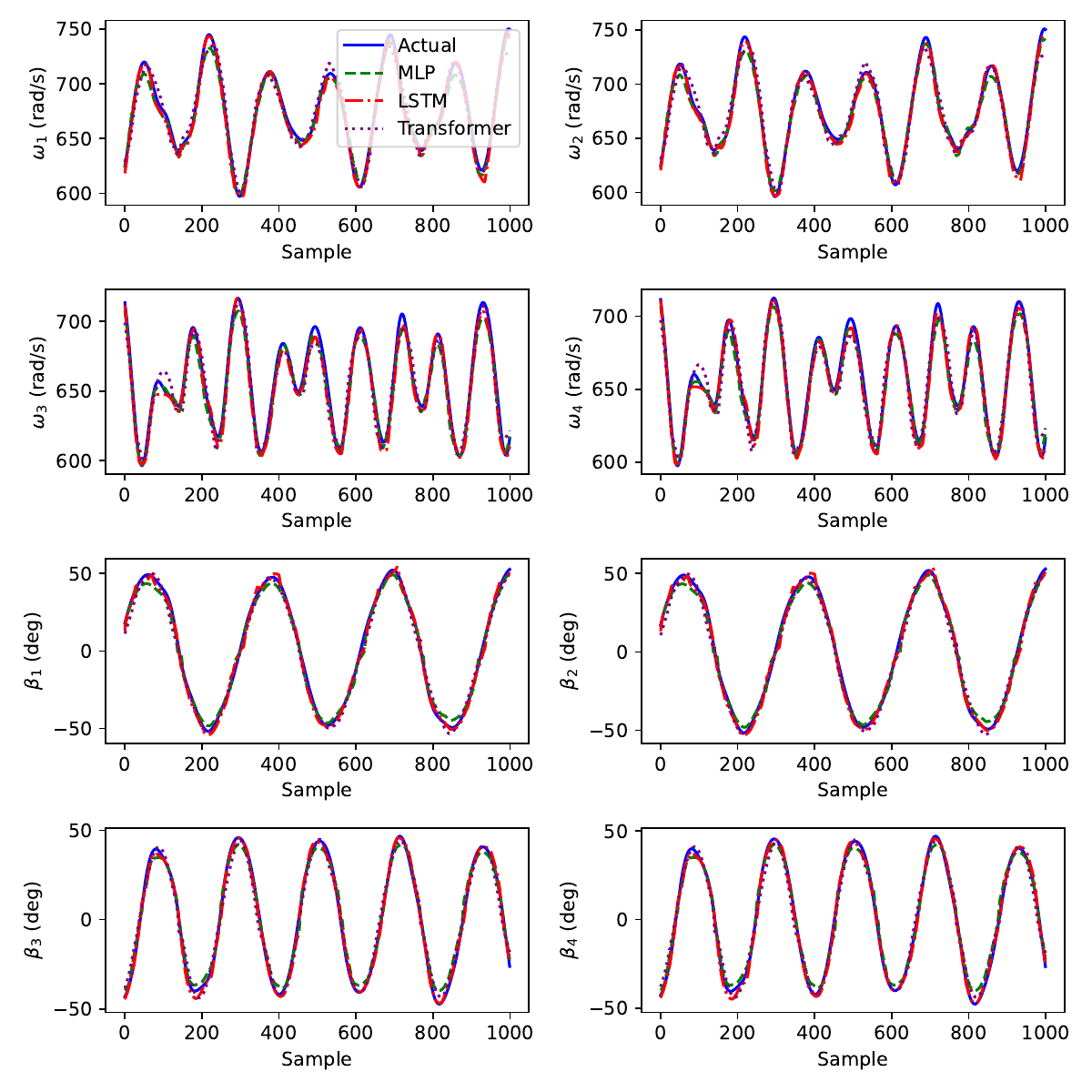}
	\caption{Prediction results of transformer-based network on 1000 train samples for direct input-output control method} 
	\label{fig:reg_train}
\end{figure}

\begin{figure}
	\centering
	\includegraphics[width=0.95\linewidth]{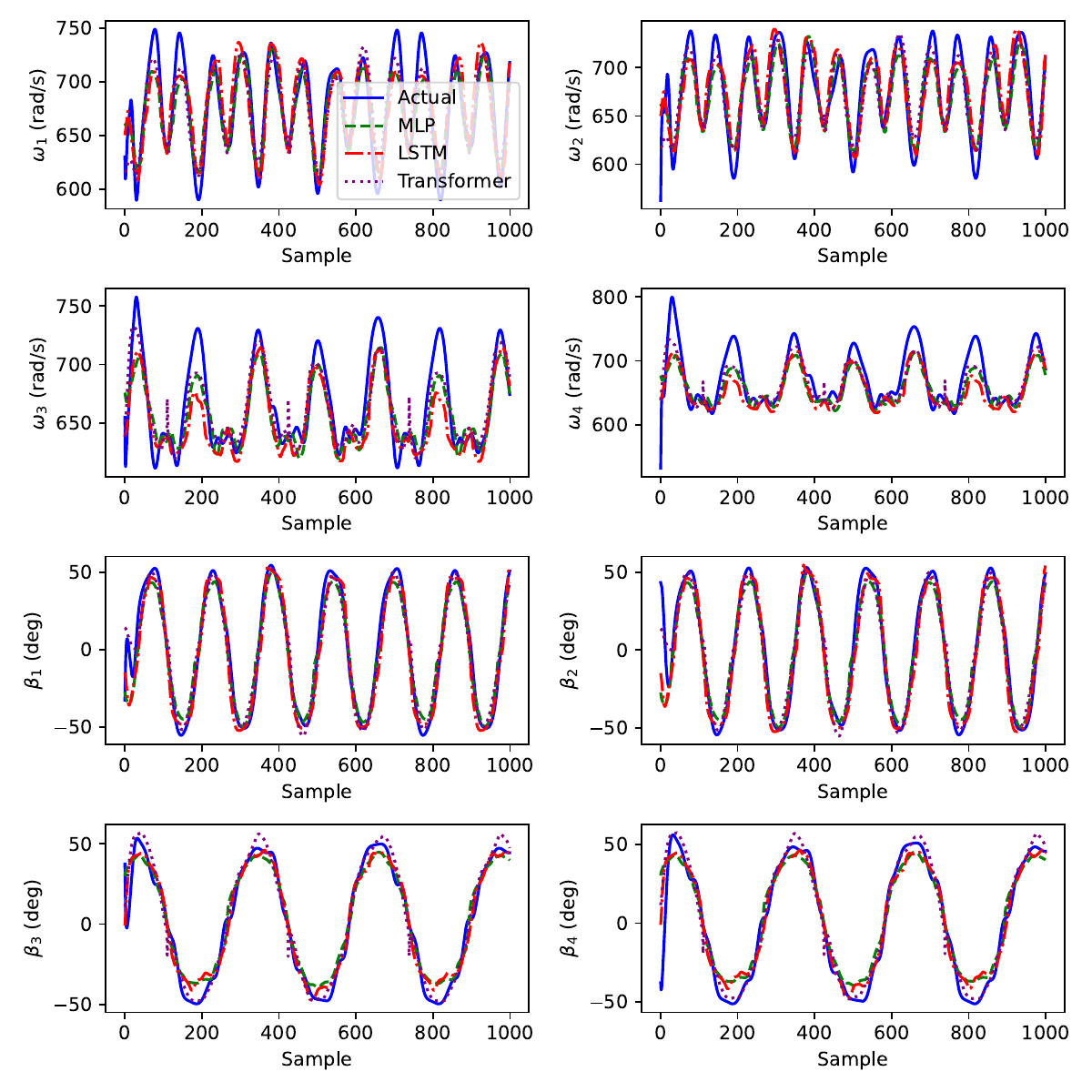}
	\caption{Prediction results on train dataset for direct input-output control method} 
	\label{fig:reg_test}
\end{figure}

The prediction results for 1000 train samples and the whole test dataset are illustrated respectively in Fig. \ref{fig:reg_train} and Fig. \ref{fig:reg_test}. Table \ref{tab:net sum dir} represents the Summary of the networks used for training direct input-output controller. As indicated by this table, the transformer network exhibited better results in predicting control signals based on state feedback and the desired state. However, overall, there is no significant difference between their results on the test set.

{\renewcommand{\arraystretch}{1.5}
	\begin{table}[h!]
		\fontsize{9pt}{9pt}\selectfont
		\centering
		\caption{Summary of the networks used for training direct input-output controller.}
		\label{tab:net sum dir}
		\begin{tabular}{| >{\centering\arraybackslash}m{1.5cm} | >{\centering\arraybackslash}m{0.8cm}| >{\centering\arraybackslash}m{1.3cm}| >{\centering\arraybackslash}m{0.75cm}|
				>{\centering\arraybackslash}m{0.75cm}|	 }
			\hline
			Network & Train Epochs & Parameters & Train MAE & Test MAE \\
			\hline
			MLP & 1400 & 45000 &  3.86 & 10.85 \\
			\hline
			LSTM & 300 & 257864 & 2.62 & 10.61 \\
			\hline
			Transformer & 10000 & 14580  & 5.26 & 9.49 \\
			\hline
		\end{tabular}
	\end{table}
}

\subsection{Implementation on the Dynamic System}
\begin{figure}
	\centering
	\begin{subfigure}{0.9\linewidth}
		\centering
		\includegraphics[width=0.98\linewidth]{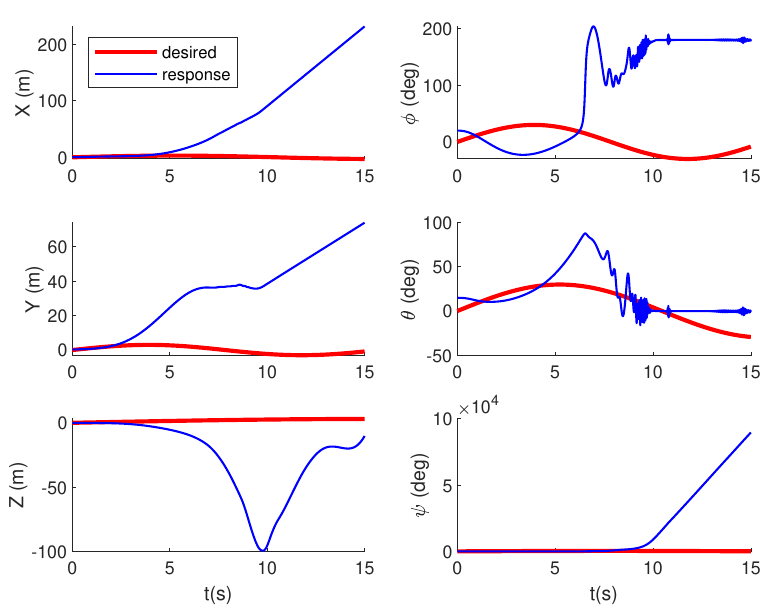}
		\caption{Responses} 
	\end{subfigure}
	\vspace{2ex}
	\begin{subfigure}{0.9\linewidth}
		\centering
		\includegraphics[width=7.5cm]{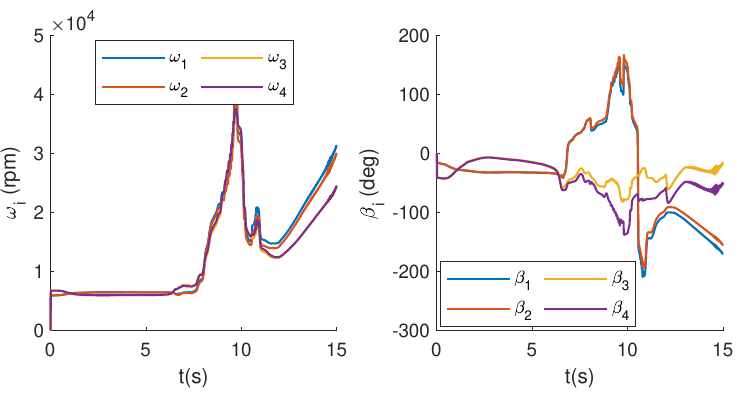}
		\caption{Control inputs} 
	\end{subfigure}
	
	\caption{Simulation results of applying the trained LSTM-based network as a controller for tracking a trained trajectory.} 
	\label{fig: LSTM-DYN}
\end{figure}

To evaluate the feasibility of the direct input-output control approach, we initially applied it to a trained trajectory using the LSTM-based network, which exhibited the best results for this purpose. The parameters for simulation are represented in Table \ref{tab:simparams}. As illustrated in Fig. \ref{fig: LSTM-DYN}, the direct input-output control method proves ineffective for our dynamic system, even with a well-fitted function as shown in Fig. \ref{fig:reg_train} for the control signal. This ineffectiveness arises because the network does not aim to minimize the error value in real-time, leading to the accumulation of small error values that ultimately cause divergence. Due to this issue, it is not feasible to compare the real-time computational cost between the conventional controller and those based on different neural networks.

\begin{table}
	\caption{Parameter values for simulations}
	\centering
	\fontsize{9pt}{9pt}\selectfont
	\label{tab:simparams}
	\begin{tabular}{|c|c|c|}
		\hline
		Parameter & Value & Unit \\
		\hline
		$g$ & $9.81$ & $m/s^2$ \\
		$m$ & $1$ & $kg$ \\
		$l$ & $0.3$ & $m$ \\
		$e$ & $0$ & $m$ \\
		$I_{xx}$ & $10 \times 10^{-3}$ & $kg.m^2$ \\
		$I_{yy}$ & $10\times10^{-3}$ & $kg.m^2$ \\
		$I_{zz}$ & $17\times10^{-3}$ & $kg.m^2$ \\
		$k_f$ & $6\times10^{-6}$ & $N.s^2$ \\
		$k_m$ & $4\times10^{-7}$ & $N.m.s^2$ \\ 
		\hline
	\end{tabular}
\end{table}

\subsection{Challenges and Limitations}
The methodology introduced in this section has several issues for our application, which will be addressed in the next section. Despite these limitations, the results obtained here can be useful for other dynamic systems or future research. The key challenges and limitations are summarized as follows:

\begin{itemize}
	\item Convergence issues even on the trained trajectories.
	\item Inability to train over a wide range of positions.
	\item Requirement of a large training dataset for accurate predictions.
	\item Lack of robustness.
	\item Inability to incorporate an effective unit to enhance control robustness due to the complex nature of the system dynamics.
\end{itemize}

\section{A Neural Network- Based Sliding Mode Control} 
\subsection{Methodology}
A precise identification of our plant dynamics is not possible by training a network using input-output pairs without a controller. This is due to the system's instability, which leads to divergence and failure within a few seconds. Therefore, in the last section, we attempted to control the system using a network without relying on its dynamics. However, we encountered some issues. If we can approximate the dynamic model of the plant, we can incorporate it into a robust model-based controller to achieve precise control. The main idea of this section is to decompose the plant's dynamics into parts that are independent of control inputs and solely dependent on the system's states, and integrate them into a \ac{SMC}. By incorporating vector $\bm{v}$ introduced in Equation \ref{v}, the approximate dynamics of the plant can be written as:

\begin{equation}
	\hat{{\ddot{\bar{\bm{x}}}}}_{6 \times 1} = \bm{\hat{a}}_{6 \times 1} + \bm{\hat{G}}_{6 \times 8} \bm{v}_{8 \times 1},
\end{equation} 

where $\bm{\hat{a}}$ and $\bm{\hat{G}}$ are the approximations of $\bm{a}$ and $\bm{G}$, respectively. $\bm{G}$ is a matrix that relates the arrangement of the actuators to the dynamics. Thus, it is evident that $\bm{G}$ varies with the attitude angles $\phi$, $\theta$, and $\psi$ considering the fixed physical parameters of the plant-\ref{tab:simparams}. Since $X$, $Y$, and $Z$ are position states expressed in the inertia frame, it is clear that $\bm{a}$ varies with $\phi$, $\dot{\phi}$, $\theta$, $\dot{\theta}$, $\psi$, and $\dot{\psi}$. Therefore, we need to predict 48 items for $\bm{\hat{G}}$ based on 3 inputs and 6 items for $\bm{\hat{a}}$ based on 6 inputs. We use the dataset represented in \ref{table:maneuvers}, but with half the number of data points for training.

Low cost computation is vital for real-time applications. Therefor, we considered to use simple architecture as possible. Also, $\bm{\hat{a}}$ and $\bm{\hat{G}}$ are not too intricate to predict, unlike the predictions discussed in the previous section. Not encountering the over-fitting issue enables the elimination of dropout layers.

\subsubsection{MLP Network}
For train and prediction of $\bm{\hat{G}}$, 2 dense layers with 128 and 64 neurons are incorporated, along with the output layer of 48 neurons. For $\bm{\hat{a}}$, 3 dense layers with 32,16, and 8 neurons are used, along with the output layer. 

\subsubsection{LSTM Network}
A LSTM block with dimension of 64 is used for $\bm{\hat{G}}$ and one with dimension of 16 is used for $\bm{\hat{a}}$.

\subsubsection{Transformer Network}
Using a Transformer for this specific application might not be appropriate choice, because they are computationally intensive and complex, adding unnecessary complexity without significant benefits for our straightforward relationships of $\bm{\hat{a}}$ and $\bm{\hat{G}}$. Additionally, Transformers require a large amount of data to train effectively. Moreover, Transformers involve more overhead in terms of model size and inference time, which might not be suitable for our real-time application, along with the other high-cost computations needed for \ac{SMC}.

\subsection{Baseline System Simulation}
\begin{figure}
	\centering
	\includegraphics[width=0.98\linewidth]{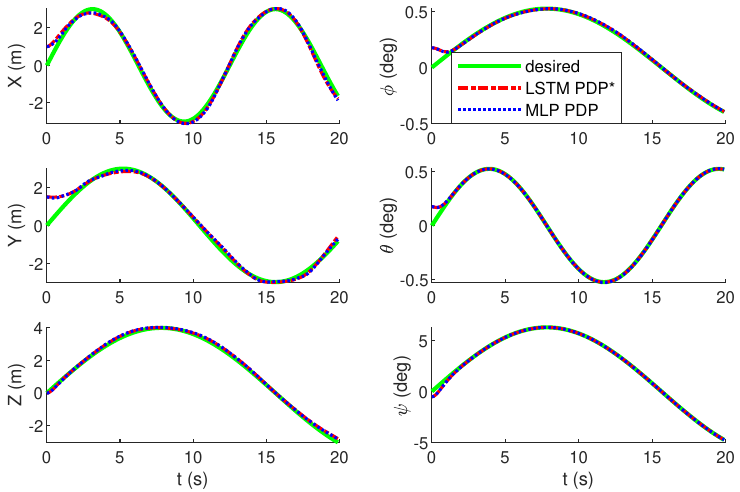}
	\caption{Baseline system simulation using neural-SMC. *PDP: Plant Dynamic Predictor} 
	\label{fig:SMC ideal}
\end{figure}

\begin{figure}
	\centering
	\begin{subfigure}{0.98\linewidth}
		\centering
		\includegraphics[width=0.98\linewidth]{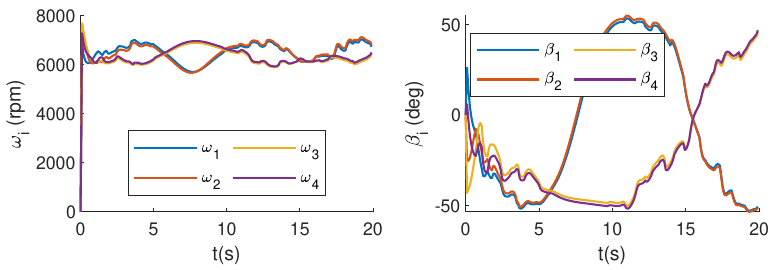}
		\caption{Using MLP plant dynamic predictor} 
	\end{subfigure}
	\vspace{2ex}
	\begin{subfigure}{0.98\linewidth}
		\centering
		\includegraphics[width=0.98\linewidth]{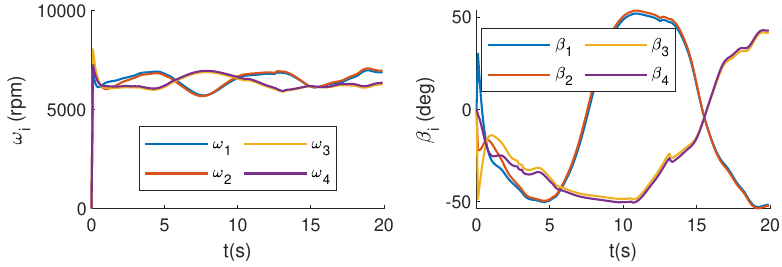}
		\caption{Using LSTM plant dynamic predictor} 
	\end{subfigure}
	
	\caption{Control signals of base-line simulation using neural-SMC.} 
	\label{fig: SMC-contrl signals-ideal}
\end{figure}

The approximations of plant dynamics parameters, $\bm{\hat{a}}$ and $\bm{\hat{G}}$, are predicted using neural networks based on necessary feedback in each loop and incorporated into the \ac{SMC} controller designed in Section \ref{sec:control}. The controller uses the predicted $\bm{\hat{a}}$ and $\bm{\hat{G}}$ in place of the unknown plant parameters $\bm{a}$ and $\bm{G}$ in Equation (\ref{Gv}). For the simulation, the controller is applied to the exact plant dynamics. The implementation results on test trajectory, using \ac{MLP} and \ac{LSTM} plant dynamics predictors are visualized in Fig. \ref{fig:SMC ideal} and Fig. \ref{fig: SMC-contrl signals-ideal}.

{\renewcommand{\arraystretch}{1.5}
	\begin{table}[h]
		\centering
		\caption{Performance comparison of MLP and LSTM networks for prediction of $\bm{a}$ and $\bm{G}$.}
		\label{tab:a G train}		
		\begin{tabular}{|c|c|c|c|c|}
			\hline
			\textbf{Net} & \textbf{Item} & \textbf{Train MAE} & \textbf{Test MAE} & \textbf{Parameters} \\ \hline
			\multirow{2}{*}{\textbf{MLP}} & $\bm{a}$ & 0.0142 & 0.0155 & 942 \\ \cline{2-5}
			& $\bm{G}$ & 0.219 & 0.222 & 11888 \\ \hline
			\multirow{2}{*}{\textbf{LSTM}} & $\bm{a}$ & 0.0049 & 0.0046 & 1574 \\ \cline{2-5}
			& $\bm{G}$ & 0.077 & 0.076 & 20528 \\ \hline
		\end{tabular}
	\end{table}
}

\subsection{Challenged System Performance and Robustness Assessment}
To assess the robustness of the control systems, we introduce various uncertainties and disturbances into the dynamic model to create discrepancies between the plant and the controller's model:

a) A 100 $\mu$s first-order delay for BLDC motors and a 0.4 degrees per 60 seconds delay for servo motors.

b) Random actuation errors ranging from 1-5 percent for servo motors and 4-10 percent for BLDC motors.

c) Variations in physical parameters: mass increased by 5\%, moments of inertia increased by 20\%, aerodynamic coefficients of propellers decreased by 10\%, and mass center eccentricity increased by 20\%.

d) Disturbance forces such as wind in the $X$ and $Y$ directions, with a random resultant force amounting to 5-10 percent of the system's weight.

\begin{figure}
	\centering
	\includegraphics[width=0.98\linewidth]{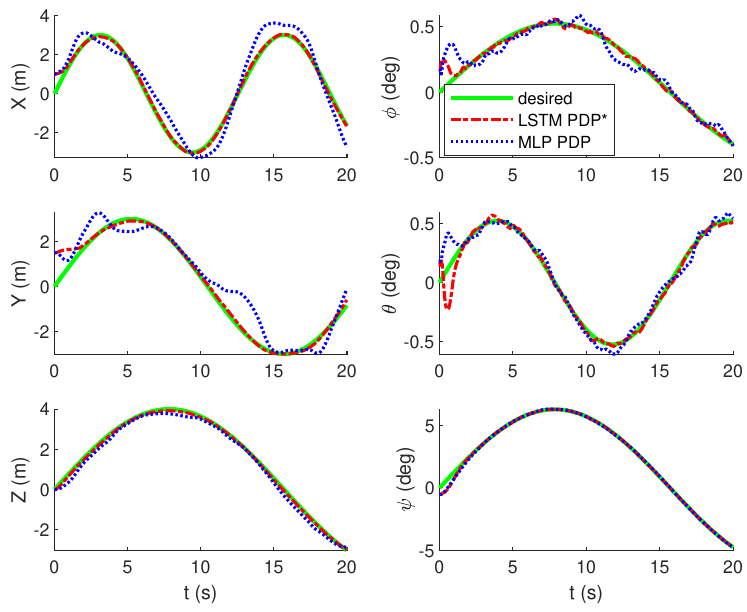}
	\caption{Simulation of the system using neural-SMC under the challenging condition (for test trajectory). *PDP: Plant Dynamics Predictor} 
	\label{fig:SMC real}
\end{figure}

\begin{figure}
	\centering
	\begin{subfigure}{0.98\linewidth}
		\centering
		\includegraphics[width=0.98\linewidth]{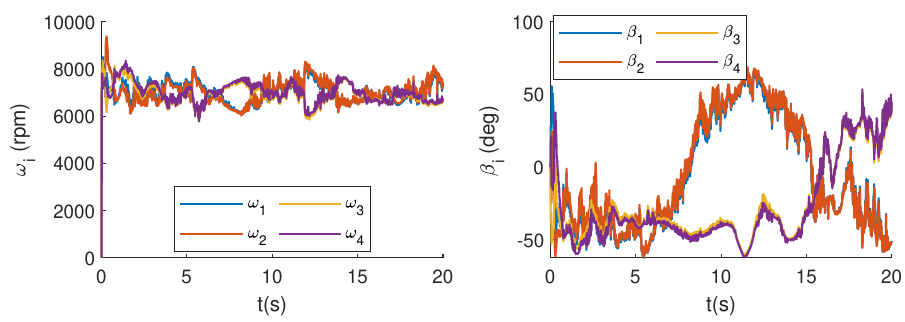}
		\caption{Using MLP plant dynamic predictor} 
	\end{subfigure}
	\vspace{2ex}
	\begin{subfigure}{0.98\linewidth}
		\centering
		\includegraphics[width=0.98\linewidth]{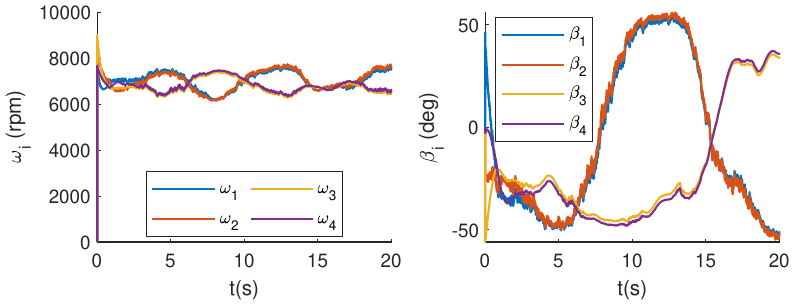}
		\caption{Using LSTM plant dynamic predictor} 
	\end{subfigure}
	
	\caption{Control signals of simulation under the challenging condition  using neural-SMC.} 
	\label{fig: SMC-contrl signals-real}
\end{figure}

Fig. \ref{fig:SMC real} and Fig. \ref{fig: SMC-contrl signals-real} illustrate the challenged system simulation results.

As shown in Figure \ref{fig:SMC real} and Table \ref{tab: SMC compare}, the control system incorporating the LSTM \ac{PDP} demonstrates superior performance compared to the controller using MLP \ac{PDP}, while maintaining nearly the same computational cost. However, the performance of the SMC-MLP system can be improved by increasing the layers of MLP, leading to a higher computational cost, which is not desirable for our real-time application.

{\renewcommand{\arraystretch}{1.5}
	\begin{table}[h]
		\centering
		\caption{Comparison of performance and computation cost of the control systems incorporating MLP and LSTM plant dynamic predictors on the test trajectory. (*PDP: Plant Dynamics Predictor, TT-MAE: Trajectory Tracking Mean Absolute Error)}
		\label{tab: SMC compare}
		\begin{tabular}{| >{\centering\arraybackslash}m{0.75cm} | >{\centering\arraybackslash}m{1.4cm}| >{\centering\arraybackslash}m{1.4cm}| >{\centering\arraybackslash}m{1.4cm}| }
			\hline
			\textbf{PDP}* & \textbf{Baseline System TT-MAE*} &  \textbf{Challenged System TT-MAE} & \textbf{Run Time (Baseline)} \\
			\hline
			\textbf{MLP} & 0.161 & 1.05 &  41.5 s  \\
			\hline
			\textbf{LSTM} & 0.156 & 0.21 & 38.5 s  \\
			\hline
		\end{tabular}
	\end{table}
}

\section{Conclusion}

This paper investigated neural-network-based control strategies for a fully actuated tilt-rotor multirotor. First, we presented a negative result by applying direct input\textendash output learning with MLP, LSTM, and transformer models, showing its inability to stabilize highly unstable dynamics.
As the main contribution, we proposed a neural-network-enhanced sliding mode controller (SMC), where input-independent dynamics are learned from a small dataset using lightweight networks. This reduces real-time computation, enables deployment on low-cost microcontrollers, and ensures robust control. The proposed method also has the advantage of being trainable using flight logs from low-performance, attitude-only controllers, without degrading the performance of the final control strategy. Comparisons of MLP- and LSTM-based implementations under uncertainties and disturbances confirmed the effectiveness of the proposed method; In particular, the controller with an LSTM plant dynamics predictor outperformed its MLP-based counterpart while also achieving lower runtime.
In summary, while direct input\textendash output strategies are unsuitable for unstable tilt-rotor systems, integrating lightweight neural networks into model-based control provides a robust, efficient, and generalizable solution. As future work, we plan to validate the outcomes of this study experimentally on an experimental tilt-rotor platform designed and constructed in our laboratory.

\bibliography{Ali_bib}
\bibliographystyle{IEEEtran}

\end{document}